\documentclass[a4paper,11pt]{article}
\usepackage{jcappub}
\usepackage[T1]{fontenc} 
\usepackage{hyperref}
\input psfig.sty
\usepackage{graphics}
\usepackage{graphicx,epsfig,youngtab}
\usepackage{epstopdf}

\def\ba{\begin{eqnarray}}
\def\ea{\end{eqnarray}}

\title{Testing the Cosmological Principle in the radio sky}

\author[a]{Carlos A. P. Bengaly,}
\author[a,b]{Roy Maartens,}
\author[a]{Nandrianina Randriamiarinarivo,}
\author[a]{{Albert Baloyi\,}}

\affiliation[a]{Department of Physics \& Astronomy, University of the Western Cape,\\
Cape Town 7535, South Africa}
\affiliation[b]{Institute of Cosmology \& Gravitation, University of Portsmouth,\\
Portsmouth PO1 3FX, United Kingdom}

\emailAdd{carlosap87@gmail.com}
\emailAdd{roy.maartens@gmail.com}
\emailAdd{nandrianiana@gmail.com}
\emailAdd{mathobela.baloyi@gmail.com}

\abstract{ 
The Cosmological Principle states that the Universe is statistically isotropic and homogeneous on large scales. In particular, this implies statistical isotropy in the galaxy distribution, after removal of a dipole anisotropy due to the observer's motion. We test this hypothesis with number count maps from the NVSS radio catalogue. We use a local variance estimator based on patches of different angular radii across the sky and compare the source count variance between and within these patches. In order to assess the statistical significance of our results, we simulate radio maps with the NVSS specifications and mask. We conclude that the NVSS data is consistent with statistical isotropy.}

\date{\today}

\keywords{}

\begin{document}
\maketitle
\flushbottom

\section{Introduction}
\label{sec:intro}

The standard $\Lambda$CDM model  provides the best current framework consistent with  observations of the cosmic microwave background (CMB) and of  the large-scale structure of the matter distribution (e.g.~\cite{Aghanim:2018eyx, Zhao:2018jxv}).  
One of the most fundamental pillars of the standard model is the Cosmological Principle, i.e., the hypothesis that the Universe follows statistical homogeneity and isotropy on large scales. Testing  these assumptions  in light of observational data is an essential robustness check of the standard cosmological model. (Note that dynamical dark energy models and modified gravity models also rely on the Cosmological Principle.) 

Statistical homogeneity is more challenging to test than statistical isotropy -- since observations can directly probe isotropy (on the observer's past lightcone), whereas spatial variations on cosmological scales (inside the observer's past lighcone) cannot be directly measured. 
Here we focus on direct tests of isotropy. 

The CMB delivers the most precise tests of isotropy, with its all-sky coverage and exquisite data. It is isotropic at a $\sim 10^{-5}$ level, after the $\sim 10^{-3}$ dipole, due to our motion relative to the CMB frame, is removed. However, near-isotropy of the CMB does {\em not} in itself imply near-isotropy of the Universe~\cite{cm2010}.  We also need independent probes of anisotropy in the matter distribution. These probes constitute a critical consistency test of statistical isotropy.

Previous work has shown no statistically significant violation of isotropy in the observational data of Type Ia Supernova distances~\cite{Huterer:2016uyq, Andrade:2018eta, Soltis:2019ryf}
and of gamma-ray bursts~\cite{Ukwatta:2015rxa, Ripa:2017scm, Andrade:2019kvl}. More stringent tests require the far higher number densities delivered by large galaxy surveys. The simplest way to test consistency with the CMB is to measure the dipole of a (sufficiently wide) galaxy survey, which should be aligned with the direction of the CMB dipole~\cite{ellis84, Baleisis:1997wx, Blake:2002gx, Crawford:2008nh, Itoh:2009vc, Singal:2011dy, Gibelyou:2012ri, Rubart:2013tx, Tiwari:2015tba, Colin:2017juj, Maartens:2017qoa, Bengaly:2017slg, Singal2019}. For currently available data sets, the matter dipole direction is not inconsistent with the CMB, but the amplitude is too large, probably arising from the quality of current data sets. Forecasts predict that future all-sky radio continuum surveys with the SKA should achieve the accuracy necessary to make a stringent test of consistency  with the CMB dipole~\cite{Bengaly:2018ykb,Pant:2018smd}.

Measuring the dipole is sufficient to reveal a possible inconsistency on the largest scales, but this test is not possible for most surveys of the matter distribution since they are not wide enough. Furthermore, in order to systematically test isotropy, as opposed to a consistency test, we need to probe smaller scales.
On smaller scales than the dipole, some isotropy tests have been performed on galaxy surveys, including infrared~\cite{Alonso:2014xca}, optical~\cite{Sarkar:2018smv} and radio~\cite{Rana:2018eut, Tiwari:2018hrs, Dolfi:2019wye, Tiwari:2019wmj} surveys. The results so far are consistent with isotropy.

In this paper, we  analyse the number counts of the widest galaxy survey, the NRAO VLA Sky Survey (NVSS). The wide  area of NVSS allows us to use a high number of sky patches for testing isotropy, increasing the statistical power of the test. In our test, we use an estimator not previously applied to galaxy surveys. 
In order to assess the statistical significance of our analysis, we produce mock data sets, using the $\Lambda$CDM background to generate  an angular power spectrum, and  using the results of simulations to estimate the clustering properties of radio sources. A log-normal code is then applied to the angular power spectrum to generate mock sky maps.    

The paper is organised as follows: section 2 describes the observational and simulated  data; section 3 discusses our estimator; section 4 presents the results and, finally, our discussion and concluding remarks are given in section 5.

\section{Data and simulations}
\label{sec:data_prep}

The data we use is from the NVSS radio continuum survey  at 1.4\,GHz~\cite{Condon:1998iy}. The NVSS catalogue covers all the northern sky, as well as most of the southern sky except for the lowest declinations, that is, $\mathrm{DEC} < 40^{\circ}$. In our analysis, we choose a flux density range following~\cite{Bengaly:2017slg}:
\begin{equation}\label{eq:S_range}
20\,\mathrm{mJy} < S < 1000 \,\mathrm{mJy} \;.
\end{equation}
This is a conservative choice in order to ensure sample completeness (lower limit) and elimination of very bright sources (upper limit). 
Since the NVSS flux densities are found to be statistically isotropic in the sky~\cite{Tiwari:2019wmj}, no further flux cutoff is performed.

We construct a mask which excises the following regions, as in~\cite{Bengaly:2017slg}:
\begin{itemize} 

\item Close to the galactic plane, i.e., $|b| \leq 10^{\circ}$.

\item Within $1^\circ$ of the local radio sources and local superclusters (see also~\cite{Colin:2017juj}).

\item Galactic foreground emission above $T=50\,$K according to the 408\,MHz continuum map in~\cite{Haslam:1982zz}.
 \end{itemize}

\noindent In addition, we follow~\cite{Blake:2002gx,Tiwari:2019wmj} and excise the region:
\begin{itemize} 
\item Within a $5^{\circ}$ degree radius around $(l,b) = (207.13^{\circ},-17.84^{\circ})$.
\end{itemize}
This region of  anomalously high source counts could bias our results. We computed the rms flux density measurement errors in the region, finding that they are compatible with the errors of the whole NVSS sample. Hence only the source counts are anomalous, not the source properties.

\begin{figure*}[!t]
\hspace*{-0.5cm}
\includegraphics[width = 5.5cm, height = 8.0cm, angle=90]{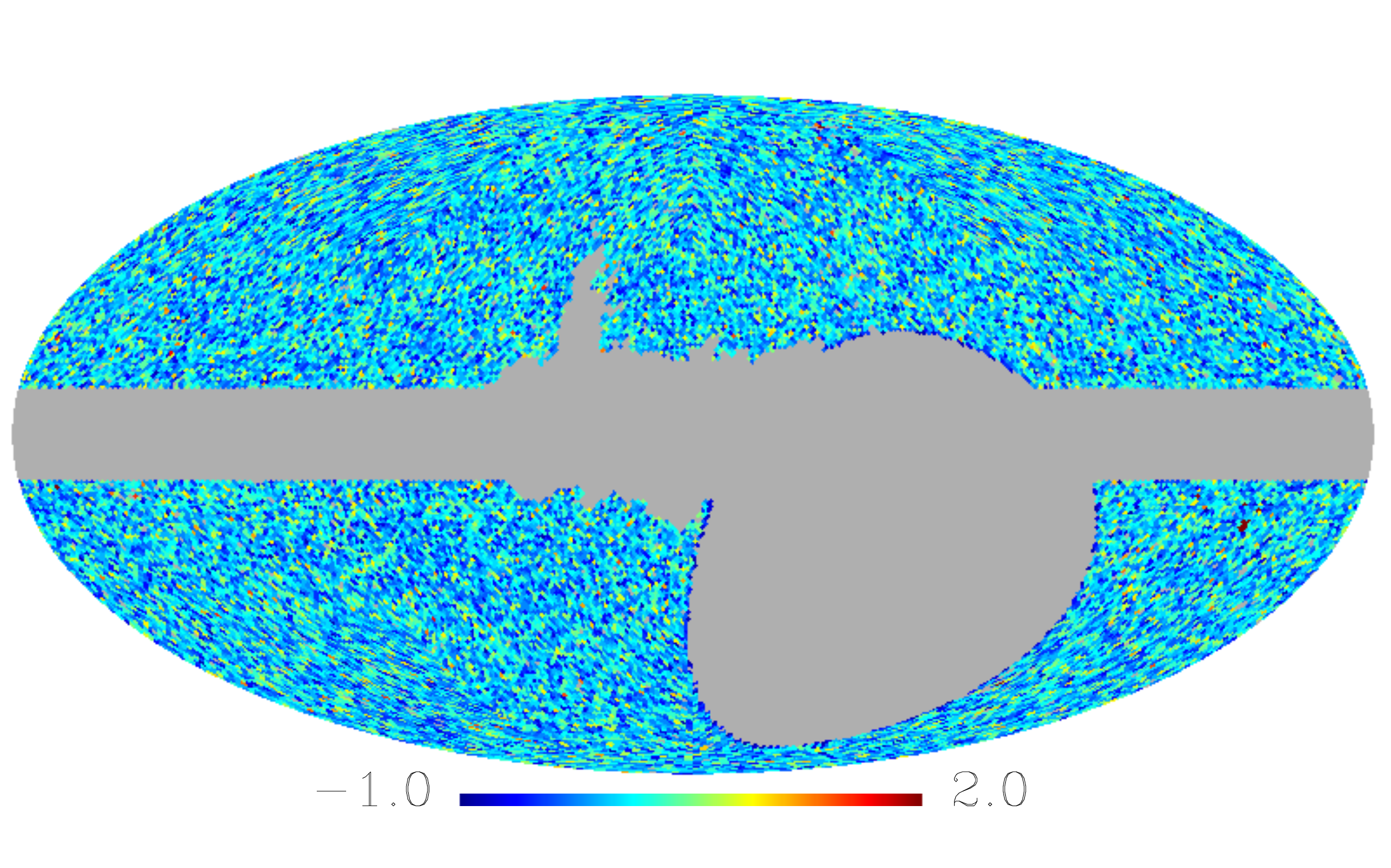}~~~
\includegraphics[width = 5.5cm, height = 8.0cm, angle=90]{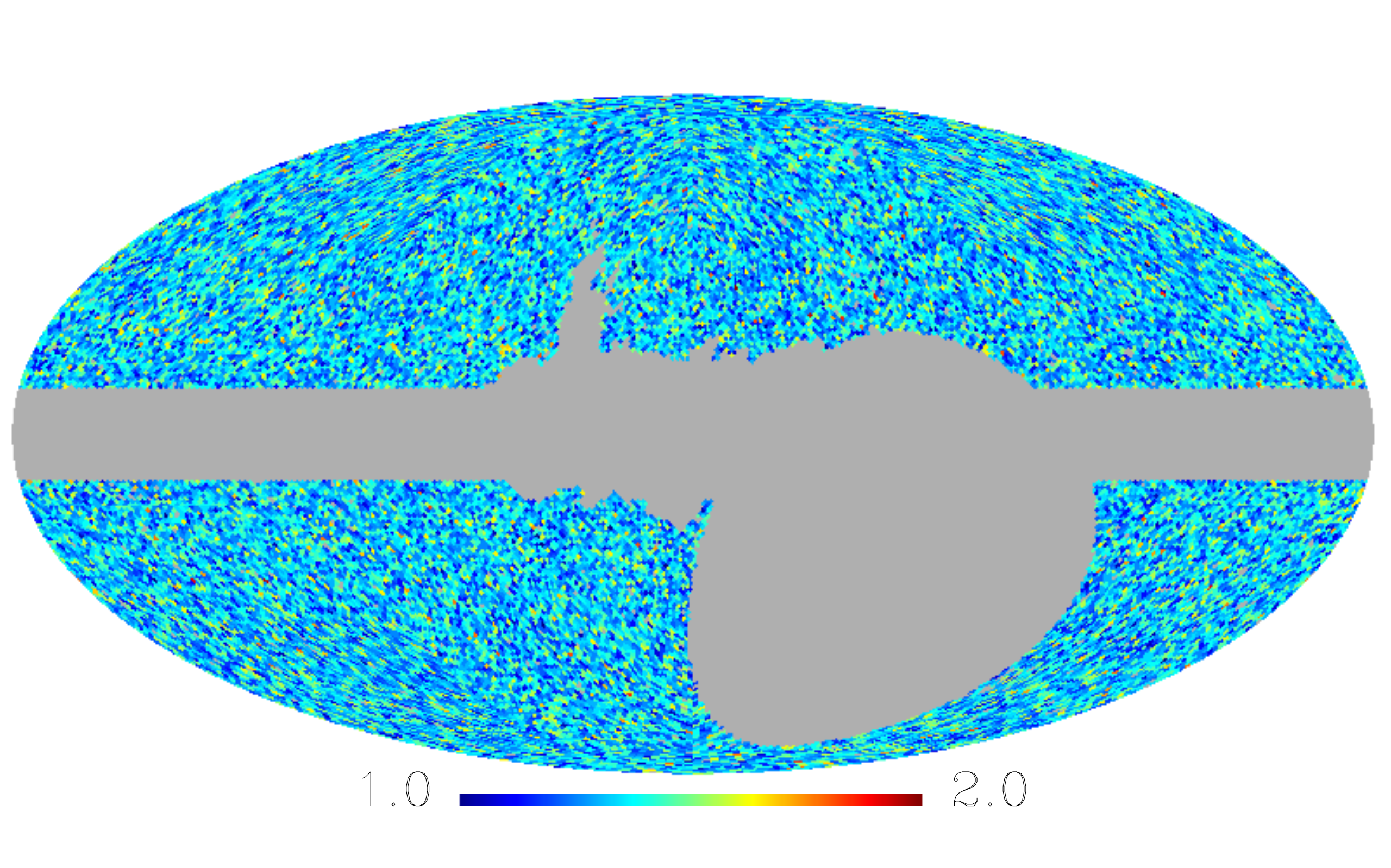}
\caption{{Real (left) and one example mock (right)} NVSS maps of number density contrast $\delta_n = (n-\bar n)/\bar{n}$, using the {\sc HEALPix} grid resolution $N_{\rm side}=64$.} 
\label{fig:nummap}
\end{figure*}

The resulting sky coverage is $f_{\rm sky} \simeq 0.657$ after producing a pixelised count map using {\sc HEALpix}~\cite{Gorski:2004by}\footnote{ \url{https://healpix.sourceforge.io/}}. Mock NVSS catalogues are produced as in~\cite{Bengaly:2017slg}: 

\begin{itemize} 

\item Choose five redshift bins in $0 < z < 4$, with edges at $z = 0.0, 0.5, 1.0, 2.0, 3.0,4.0$.
 
\item Obtain the redshift distribution of  radio sources from an {\sc SQL} query over 121\,deg$^2$ 

on the $S^3$ (SKA Simulated Skies)~\cite{Wilman:2008ew} website\footnote{ \url{http://s-cubed.physics.ox.ac.uk/s3_sex}}.

\item Model the redshift-dependent bias following~\cite{Tiwari:2015tba}:  $b(z) = 1.6  + 0.7z + 0.35z^2$.

\item Compute the angular power spectrum  using {\sc CAMB Sources}~\cite{Challinor:2011bk} in each  redshift bin, using the Planck 2015 $\Lambda$CDM best-fit parameters~\cite{Ade:2015xua}. (Note that the $\Lambda$CDM best-fit power spectrum changes very little between Planck 2015 and 2018.)

\item In each redshift bin, input the CAMB angular power spectrum into the {\sc FLASK} code~\cite{Xavier:2016elr} to produce 1000 NVSS realisations, following a lognormal distribution for density fluctuations. Produce mock maps in each bin, stacking them afterwards to produce a full NVSS map covering $0 < z < 4$.

\end{itemize}

This prescription generates mock realisations that are statistically isotropic. Then we add to the mock counts a fiducial kinematic dipole signal, based on the measurement reported in~\cite{Bengaly:2017slg}  (see~\cite{Bengaly:2018ykb} for details).
The pixelised count maps of the real data, and one of the NVSS realisations, are shown in Figure~\ref{fig:nummap}.

\section{Probing isotropy}
\label{sec:est}

We perform a statistical isotropy test on the NVSS data according to the following prescription.

\begin{itemize} 

\item We define 3072 directions in the sky coinciding with the $N_{\rm side} = 16$ pixel centres. 

\item Around each centre, we draw a patch of angular radius $\theta$, where 
\ba \label{theta}
\mbox{angular size of patches:}\qquad \theta=15^{\circ},20^{\circ},25^{\circ}~\mbox{and}~30^{\circ}\,.
\ea

\item When a patch overlaps the  masked sky area, we need to decide a threshold of masking in the patch, above which the patch should be rejected for the test (see also~\cite{Akrami:2014eta, Alonso:2014xca}). 
If $m$ is the threshold percentage of masked pixels in a patch, then we reject the patch if its percentage of masked pixels is $>m$. We choose  
\ba \label{m}
\mbox{threshold of masked pixels in patch:}\qquad m = 10\%, 30\%~\mbox{and}~50\%\,,
\ea
i.e., from strict to weak rejection criteria. Note that our criteria is stricter than those assumed in~\cite{Akrami:2014eta, Alonso:2014xca}, which used $m=90\%$, in order to avoid further biases due to incomplete sky coverage.

\end{itemize}

A patch $p$ has $\mu_p$ pixels, with $n_{pi}$ radio sources in pixel $i$. The total number of sources in the patch is  $n_p=\sum_i n_{pi}$, and the average number per pixel is  $\bar n_p=n_p/\mu_p$.
Then the variance in the patch $p$ is 
\begin{equation}
\sigma_{p}^2 = \frac{1}{\big(\mu_p - 1\big)}\,\sum\limits_{i=1}^{\mu_p} \big(n_{pi} - \bar n_p\big)^2\,.
\label{eq:var}
\end{equation}

We assess the level of statistical isotropy of the NVSS sample in different patches by means of the AnoVa (Analysis of Variance) approach~\cite{Gravetter2012}, which  is a collection of methods to compare multiple means. We use it to produce a comparison of the mean number counts between the patches, and also within each patch.  

Like most statistical tests, we accept or reject the null hypothesis of a certain feature of the data. AnoVa uses the $F$-ratio to provide a quantity that measures the differences amongst all of the mean  counts  in patches~\cite{Gravetter2012}. It is defined as: 
\begin{equation}
F = \frac{ \mathrm{variance\  between\  patches}}{\mathrm{variance\  within\  patches}}\,.
\end{equation}
Then we have
\ba\label{f}
\mbox{exactly isotropic data:}\qquad F= 1\,.
\ea
Since the real and mock counts are generated by gravitational clustering, we expect some deviation from 1 even for statistically isotropic data. 

We need to define a criterion by which we accept or reject the null hypothesis, that is, that the Universe is not statistically isotropic. We use the 5th and 95th percentiles of the $F$-distribution as the upper limit of statistical isotropy -- noting that the mock realisations are statistically isotropic by construction. If the $F$-value of the real data disagrees with that, it would be an indication of statistical isotropy violation. If not, we reject the null hypothesis.

In addition to the AnoVa test, we use another method to assess the statistical isotropy of the NVSS data. It is based on the Local Variance (LV) estimator defined in~\cite{Akrami:2014eta} (see also~\cite{Alonso:2014xca}), and it provides a more direct method to visualise the deviation of the mean counts per patch across the sky between the real and the mock data. The LV estimator is defined for each patch as
\begin{equation}
\label{eq:lv_eq}
{\rm LV}_p = \frac{\big({\sigma_p}/{M}\big)_{\rm data} - {\big(\overline{{\sigma_p}/{M}}\big)}_{\rm mock}}{\big(\overline{{\sigma_p}/{M}}\big)_{\rm mock}} \;,
\end{equation}
where $\sigma_p$ is given by \eqref{eq:var} and $M$ is the average source count per patch. This is given by $M=\sum_p n_p/N$, where $N$ is the number of accepted patches. We compare the coefficient of variation in every patch drawn in the real data map, $({\sigma_p}/{M})_{\rm data}$, with the average coefficient  of variation per patch over all  mock realisations, $\big(\overline{{\sigma_p}/{M}}\big)_{\rm mock}$.

Note that the LV estimator only provides visual information of the variance fluctuation across the sky, which is why we deploy the $F$-ratio and $F$-distribution  as a metric to quantify how consistent with the statistical isotropy assumption the NVSS data is.     

\section{Results }
\label{sec:res_dis}

\begin{figure}[t!]
\hspace*{-3.2cm}
\includegraphics[scale = 0.45]{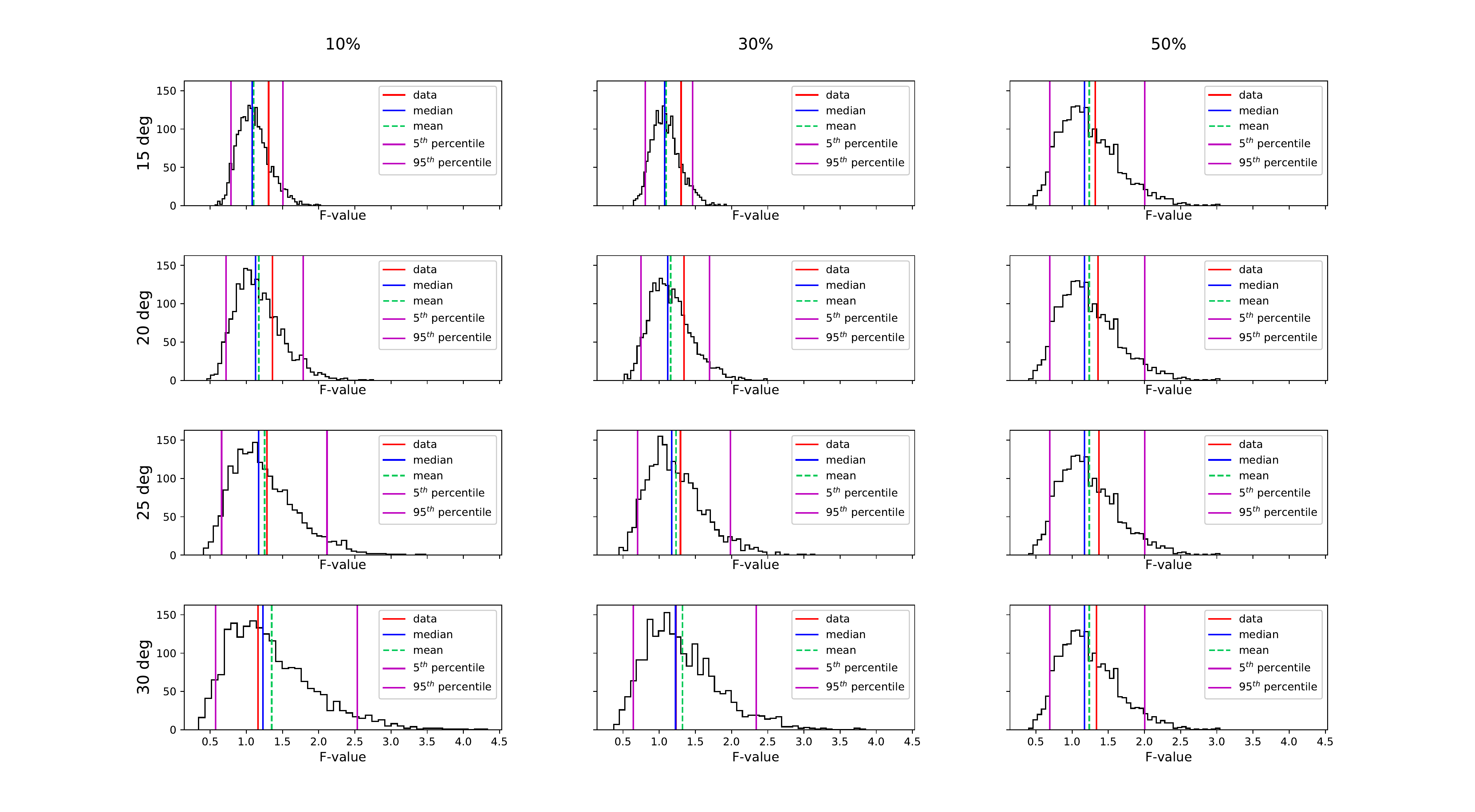}
\caption{$F$-distribution of the NVSS realisations (black histogram) and  $F$-value of the NVSS data (red vertical line), for 4 patch radii  (rows) and 3 masked pixel rejection thresholds (columns).  Also shown are: 5th and 95th percentiles (magenta), the mean (dashed green) and the median (blue) of the $F$-distribution.}
\label{fig:fr_new}
\end{figure}
 
\begin{figure}[h!]
\begin{center}
\hspace*{-1.8cm}
\includegraphics[scale = 0.38]{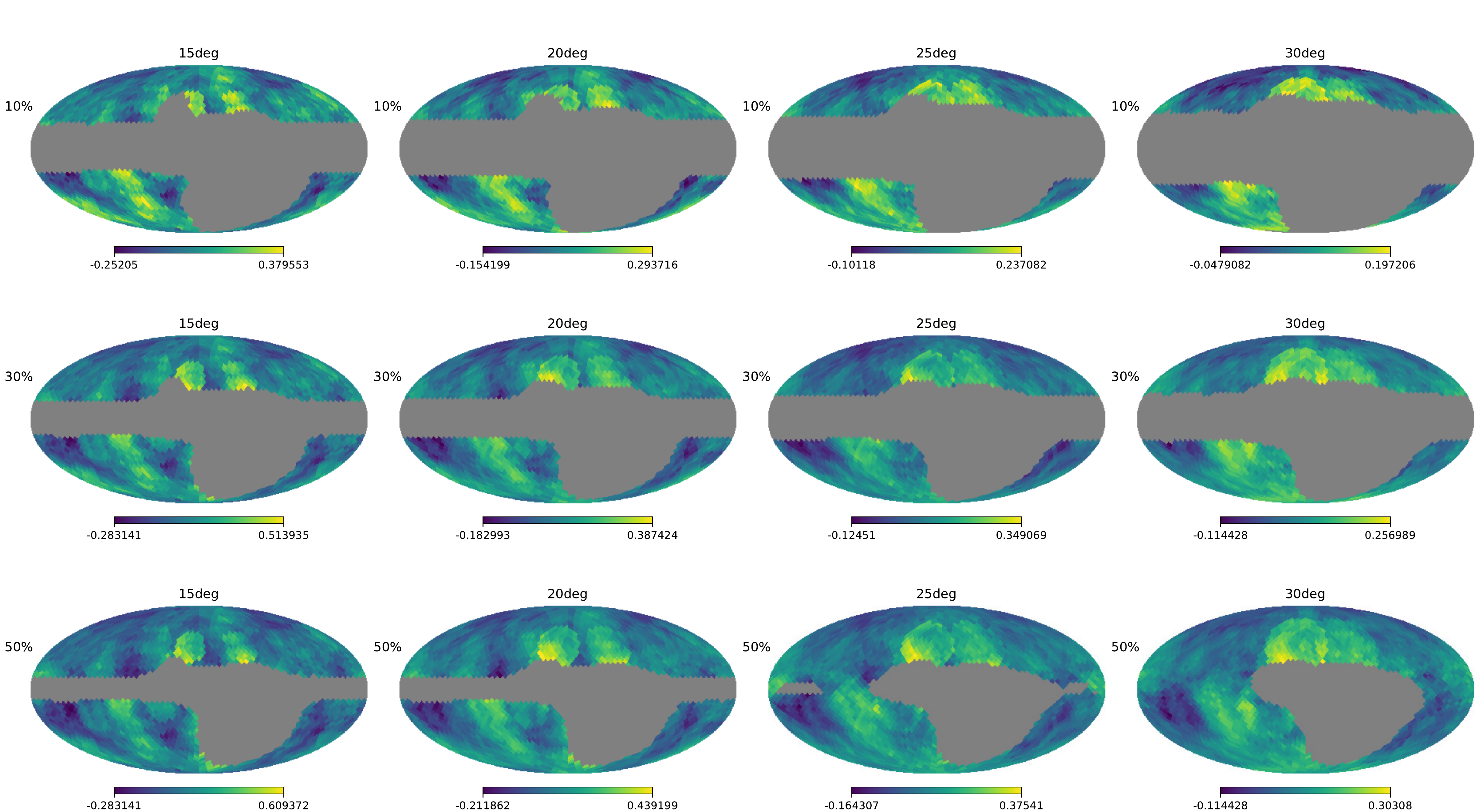}
\caption{Map of local variance \eqref{eq:lv_eq}, for 4 patch radii  (columns) and 3 masked pixel rejection thresholds (rows).}
\label{fig:LV_new}
\end{center}
\end{figure}

We show the $F$-distribution of the mock realisations as compared to the real NVSS data in Figure~\ref{fig:fr_new}. 
All the distributions peak around 1. Vertical lines indicate the 5th and 95th percentiles, the mean and the median of the $F$-distribution, together with the $F$-statistic of the real data. 

The $F$-distribution exhibits a larger tail-end at $F>1$ as we increase the patch radius. This is an expected feature, since the shape of the distribution depends on the degrees of freedom, and  we have a smaller number of accepted patches for larger patches. 

We note that the $F$-value obtained from the real data is in excellent agreement with both the mean and the median $F$-values of the mock realisations, regardless of the patch radius size and the rejection criterion adopted. They are all close to the vertical line around 1, and well within the 5th and 95th boundaries obtained from the simulations.

Statistical isotropy of the source counts is also evident in the local variance maps shown in Figure~\ref{fig:LV_new}: there is no indication of preferred directions, or anomalous variance, across the sky. 

Thus, on the basis of the $F$  and local variance tests, we can safely reject the null hypothesis -- that the Universe is not statistically isotropic -- and conclude that the NVSS catalogue is indeed statistically isotropic at $15^{\circ}-30^{\circ}$ angular scales.

\section{Conclusions}

In this work, we tested whether the NVSS catalogue is consistent with the statistical isotropy hypothesis on angular scales smaller than the dipole -- whose amplitude is known to be larger than expected. 
We constructed a mask that extends the mask in~\cite{Bengaly:2017slg} by excising a small region of anamolously high source counts, following~\cite{Blake:2002gx,Tiwari:2019wmj}. 
We simulated radio sky maps that reproduce the NVSS specifications, as well as the radio source clustering and the power spectrum of the standard model, following~\cite{Bengaly:2017slg}. A dipole modulation consistent with the kinematic dipole signal observed in the real data was also applied in these mocks~\cite{Bengaly:2017slg,Bengaly:2018ykb}. 

Our test consists of drawing patches across the sky with different angular radii, \eqref{theta}, and different rejection thresholds  to eliminate masked pixels within the patches, \eqref{m}. We apply an LV estimator, in addition to the AnoVa test, to compare the data with the simulations. This AnoVa test provides us with an $F$-value for the real data, and a distribution of $F$-values  for the mock realisations, as a metric that quantifies how consistent are these patches across the sky. In other words, it quantifies how isotropic these source count maps are. Because the mocks are statistically isotropic by construction, they  constitute a benchmark for isotropy. Hence, our null hypothesis is that the real data is not consistent with the mocks -- i.e., that the NVSS is not consistent with statistical isotropy -- within the 5th--95th percentile region of the $F$-distribution.

We found that the observed radio counts reject the null hypothesis, since the data agrees with the mock maps produced for patches of all chosen radii and with all masked pixel rejection thresholds. As we increase the patch radius, the discrepancy between the data and simulations becomes smaller, since we encompass a larger number of sources so that the shot noise is reduced. In addition, we noticed that the more rigorous we were in the rejection of masked pixels, the better is the agreement between the real data and statistically isotropic realisations. We found an optimum choice for patch radius of  $25^{\circ}$ and for rejection threshold of 30\%.

Therefore, there is no signal of isotropy violation in the NVSS source counts on smaller angular scales than the dipole.
This result agrees with the findings of~\cite{Tiwari:2018hrs,Dolfi:2019wye,Tiwari:2019wmj}, using other methods of analysis. A more thorough investigation of the impact of clustering bias models (which can affect the large angular scales), as well as of magnification bias and redshift distribution modelling of the source counts, is left for future work. 

Finally, we stress that our method is completely general, and can be readily applied to  larger and more complete data sets from next-generation surveys,  such as forthcoming radio continuum surveys with the SKA precursor survey EMU~\cite{Norris:2011ai} and with SKA Phase 1~\cite{Bacon:2018dui}, and future redshift surveys with SKA1~\cite{Bacon:2018dui},  J-PAS~\cite{Benitez:2014ibt}, LSST~\cite{Abell:2009aa} and Euclid~\cite{Amendola:2016saw}.

\newpage
\[\]{\bf Acknowledgements:}\\
We thank Matt Jarvis and Mario Santos for useful discussions. This work was supported by the South African SKA Project and the National Research Foundation of South Africa (Grant No. 75415). RM was also supported by the UK STFC (Grant ST/S000550/1). Some of our results made use of the {\sc HEALpix} package~\cite{Gorski:2004by}.

\end{document}